\documentclass[12pt]{iopart}
\usepackage{graphicx}
\usepackage{multirow}
\usepackage{cite}

\begin{document}
\title[On three-dimensional rotational averages of odd-rank tensors]{On three-dimensional rotational averages of odd-rank tensors}
\author{Tuguldur~Kh.~Begzjav$^1$, Reed~Nessler$^{1,2}$,
Marlan~O.~Scully$^{1,2,3}$ and Girish~S.~Agarwal$^{1,4}$}
\address{$^1$ Institute for Quantum Science and Engineering, Department of Physics and Astronomy, Texas A\&M University, College Station, TX 77840, USA}
\address{$^2$ Department of Physics, Baylor University, Waco, TX 76706, USA}
\address{$^3$ Department of Mechanical and Aerospace Engineering, Princeton University, Princeton, NJ 08544, USA}
\address{$^4$ Department of Biological and Agricultural Engineering, Texas A\&M University, College Station, TX 77843, USA}
\ead{mn.tuguldur@tamu.edu}
\begin{abstract}
The recent growing interest in nonlinear optical spectroscopy in optically active medium demands the three-dimensional rotational average of high-rank tensors.
In the present paper, we present a new method for finding the rotational average of odd-rank tensors in an overcomplete basis of isotropic tensors. The method is successfully applied to the rotational averages of tensors of rank $n=5,7,9,11$. 
\end{abstract}
\noindent{\it Keywords\/}: nonlinear spectroscopy, isotropic tensor, ninth-rank tensor, rotational average\\
\submitto{\jpa}

\section{Introduction}
\label{intro}
In most nonlinear optical problems, we work in a lab-fixed frame of reference, but the molecules comprising the system are oriented randomly with respect to that frame \cite{Craig1998}. In this situation, averaging molecular quantities over the random orientation of the molecules is usually of great interest. Moreover, the three-dimensional rotational average of high-rank isotropic tensors often appears in the theory of nonlinear spectroscopy and has been extensively examined in the physical and mathematical context in the last half century \cite{Smith1968,Boyle1970,Boyle1971,Andrews1977,Andrews1981,Wagniere1982,Andrews1989,Smith2011,Man2014,Friese2014,Ford2018}. For example, coherent anti-Stokes Raman scattering in optically active medium \cite{Bjarnason1980,Oudar1982} is a four-photon process that requires ninth-rank tensor averaging. Recently, this problem is receiving renewed interest as the demand for developing nonlinear spectroscopic tools in optically active medium increases \cite{Koroteev1995,Zheltikov1999,Hiramatsu2015,Begzjav2018}.

Let $T_{\lambda_1\lambda_2\cdots\lambda_n}$ be a tensor of rank $n$, where the indices $\lambda_1,\lambda_2,\ldots,\lambda_n$ refer to coordinates in the molecule-fixed frame. Then the tensor $T$ in the lab-fixed frame turns out to be
\begin{equation}
T_{i_1i_2\cdots i_n}=l_{i_1\lambda_1}l_{i_2\lambda_2}\cdots l_{i_n\lambda_n}T_{\lambda_1\lambda_2\cdots\lambda_n}
\end{equation} 
where $i_1,i_2,\ldots,i_n$ are coordinates in the lab-fixed frame. Here $l_{i_1\lambda_1}, l_{i_2\lambda_2}, \ldots, l_{i_n\lambda_n}$ denote direction cosines that
can be expressed in terms of Euler angles, so that a straightforward method for computing the rotational average of the tensor $T$ is an integral over Euler angles:
\begin{equation}\label{average0}
\langle T_{i_1i_2\cdots i_n}\rangle=\left(\frac{1}{8\pi^2}\int_{0}^{2\pi} \mathrm{d}\psi \int_{0}^{2\pi} \mathrm{d}\phi \int_{0}^{\pi}\sin\theta\,\mathrm{d}\theta\, l_{i_1\lambda_1}l_{i_2\lambda_2}\cdots l_{i_n\lambda_n}\right) T_{\lambda_1\lambda_2\cdots\lambda_n}.
\end{equation}
The expression in parentheses in Equation \ref{average0} is a rotational average of direction cosines and denoted by $I^{(n)}_{i_1i_2\cdots i_n;\lambda_1\lambda_2\cdots\lambda_n}$ i.e.
\begin{equation}\label{integration}
I^{(n)}_{i_1i_2\cdots i_n;\lambda_1\lambda_2\cdots\lambda_n}=
\frac{1}{8\pi^2}\int_{0}^{2\pi} \mathrm{d}\psi \int_{0}^{2\pi} \mathrm{d}\phi \int_{0}^{\pi} \sin\theta\,\mathrm{d}\theta\,l_{i_1\lambda_1}l_{i_2\lambda_2}\cdots l_{i_n\lambda_n}.
\end{equation}
This integral can be evaluated easily in the case of low-rank tensors but for higher-rank tensors ($n > 4$) its evaluation requires prohibitively much labor or computer time, in general.

If $f_r^{(n)}$ and $g_\alpha^{(n)}$ are the $r$th and $\alpha$th linearly independent isotropic tensors of rank $n$ in lab- and molecule-fixed frames, respectively then, according to Weyl's theorem \cite{Weyl1939book}, the rotational average of direction cosines $I^{(n)}_{i_1\cdots i_n;\lambda_1\cdots\lambda_n}$ can be uniquely expressed as a linear combination of products of the tensors $f_r^{(n)}$ and $g_\alpha^{(n)}$. Explicitly,
\begin{equation}\label{main0}
I^{(n)}_{i_1\cdots i_n;\lambda_1\cdots\lambda_n}=\sum_{r,\alpha}
M_{r\alpha}^{(n)}f_r^{(n)}g_\alpha^{(n)}
\end{equation}
where coefficients are denoted by $M_{r\alpha}^{(n)}$.
Therefore, first of all, it is essential to establish complete bases $f_r^{(n)}$ and $g_\alpha^{(n)}$, and second, to find the matrix $\mathbf{M}^{(n)}$ of coefficients $M_{r\alpha}^{(n)}$.

Now we state some properties of isotropic tensors of rank $n$. For even rank $n$, any product of $n/2$ Kronecker deltas is isotropic, whereas for odd rank $n$, any product of one Levi-Civita epsilon tensor and $(n-3)/2$ Kronecker deltas is isotropic. For example, $\delta_{i_1i_8}\delta_{i_2i_7}\delta_{i_3i_6}\delta_{i_4i_5}$ and $\epsilon_{i_1i_3i_5}\delta_{i_2i_6}\delta_{i_4i_8}\delta_{i_7i_9}$ are isotropic tensors of rank $8$ and $9$. By simply permuting all indices in products one can find a full (i.e.\ spanning) set of isotropic tensors of a given rank $n$ whose number is given by
\begin{equation}
\label{cases}
N_n=\cases{\frac{n!}{2^{n/2}(n/2)!} & for even $n$\\
\frac{n!}{3\cdot 2^{(n-1)/2}((n-3)/2)!} & for odd $n$\\}
\end{equation}
These isotropic tensors are not linearly independent in general. A method to find a linearly independent subset of the full set of $N_n$ isotropic tensors was developed by G.~F.~Smith \cite{Smith1968} using standard Young tableaux in 1968. Using this linearly independent set of isotropic tensors one can easily find the rotational average $I^{(n)}$ (i.e. $\mathbf{M}^{(n)}$).

However, the full set of $N_n$ isotropic tensors is more convenient for expressing the rotational average of direction cosines, especially for odd-rank tensors. Therefore, we develop a new method that allows us to find the rotational average $I^{(n)}$ for odd-rank tensors in the linearly dependent set of isotropic tensors. We use a prime sign in the matrix, $\mathbf{M}'^{(n)}$, to indicate that it is with respect to the linearly dependent set or overcomplete basis.

\section{Method}
\label{sec:1}
We begin with Equation~\ref{main0} in the overcomplete isotropic tensor basis:
\begin{equation}\label{main11}
I^{(n)}_{i_1\cdots i_n;\lambda_1\cdots\lambda_n}=\sum_{r,\alpha}
M_{r\alpha}'^{(n)}f_r^{(n)}g_\alpha^{(n)}.
\end{equation}
Fortunately, the matrix $M_{r\alpha}'^{(n)}$ turns out to depend only on a small number of independent coefficients, much fewer than the size $N_n\times N_n$ of the full matrix. We aim to find these independent coefficients, and in order to achieve this goal, we first analyze the structure of Equation~\ref{main11} and then select linearly independent equations sufficient to determine the independent coefficients in the $M_{r\alpha}'^{(n)}$.


For odd rank $n$, isotropic tensors can be classified into groups, each corresponding to a unique epsilon tensor and its members distinguished only by Kronecker deltas; for example, $\epsilon_{i_1i_2i_3}f^{(n-3)}_r$, $\epsilon_{i_1i_2i_4}f^{(n-3)}_r$ and so on. Here, $f^{(n-3)}_r$ runs over the full set of isotropic tensors of rank $n-3$. 
The matrix $\mathbf{M}'^{(n)}$ for odd rank has a block diagonal form, and each block has the same structure as $\mathbf{M}'^{(n-3)}$ \cite{Andrews1977}.
Thus, the number of independent coefficients in $\mathbf{M}'^{(n-3)}$ equals the number of independent coefficients in $\mathbf{M}'^{(n)}$.
The common value, denoted by $c_n$ for odd $n$, is given by the partition function $\mathrm{p}((n-3)/2,3)$, which counts the number of partitions of $(n-3)/2$ into at most 3 parts \cite{Andrews1977}. For example, $c_n=1,2,3,4$ for $n=5,7,9,11$.
We enumerate these independent coefficients $a,b,c,\ldots$ using the first $c_n$ letters of the alphabet.

On the left-hand side of Equation~\ref{main11} each of the $2n$ indices can take the values $x,y,z$, so the tensor $I^{(n)}_{i_1\cdots i_n;\lambda_1\cdots\lambda_n}$ has $3^{2n}$ components. Therefore, Equation~\ref{main11} can be understood as a set of $3^{2n}$ linear equations in the $c_n\ll3^{2n}$ variables $a,b,\ldots$. By definition, Equation~\ref{main11} always has the same number of linearly independent equations as the number of independent coefficients. Moreover, Equation~\ref{main11} is overcomplete and there are many linearly dependent equations. A practical question arises: how to shrink the overcomplete set of equations into a \emph{minimal} set of equations by selecting a linearly independent subset?

Here, we make the brave assumption that the diagonal terms of $I^{(n)}_{i_1\cdots i_n;\lambda_1\cdots\lambda_n}$ suffice to produce equations determining the independent coefficients. By ``diagonal terms" we mean that the indices satisfy $i_1=\lambda_1$, $i_2=\lambda_2$, $\ldots$, $i_n=\lambda_n$. The number of such terms is $3^n>c_n$, so the corresponding equations
\begin{equation}\label{main2}
I^{(n)}_{i_1\cdots i_n;i_1\cdots i_n}=\sum_{r,\alpha}
M_{r\alpha}'^{(n)}f_r^{(n)}g_\alpha^{(n)}
\end{equation}
are still overcomplete for the independent coefficients.

To pare down Equations~\ref{main2} a final time into a minimal subset, we analyze both sides of the equations in turn. 
\begin{enumerate}
\item Left-hand side: for convenience, we denote the diagonal terms $I^{(n)}_{i_1\cdots i_n;i_1\cdots i_n}$ by $I(q,r,s)$ in the manner
\begin{equation}
I(q,r,s)=\langle l_{xx}^q l_{yy}^r l_{zz}^s\rangle,
\end{equation}
i.e.\ by collecting the indices as
\begin{equation}
i_1\cdots i_n=\underbrace{x\cdots x}_{\mbox{{\it q} times}}\underbrace{y\cdots y}_{\mbox{{\it r} times}}\underbrace{z\cdots z}_{\mbox{{\it s} times}},
\end{equation}

where $q+r+s=n$. It is important to see that $I(q,r,s)$ is invariant under the permutation of indices $x$, $y$ and $z$, or equivalently the permutation of the powers $q$, $r$ and $s$. For example, we can swap $y$ and $z$ (equivalently $r$ and $s$) by simultaneously rotating the lab-fixed and molecule-fixed frames using the $90^\circ$ rotation matrix
\begin{equation}\label{matrixR}
R=\pmatrix{
-1 & 0 & 0\cr
0 & 0 & 1\cr
0 & 1 & 0
}.
\end{equation}
We have $(R l R)_{xx}=l_{xx}$, $(R l R)_{yy}=l_{zz}$ and $(R l R)_{zz}=l_{yy}$. Then from the rotational invariance of $I^{(n)}$ we obtain the desired property: 
\begin{equation}
\langle l_{xx}^q l_{yy}^r l_{zz}^s\rangle=
\langle l_{xx}^q l_{zz}^r l_{yy}^s\rangle=
\langle l_{xx}^q l_{yy}^s l_{zz}^r\rangle.
\end{equation}
The proof for other indices and powers is straightforward.

The essential outcome of this invariance property is that the components of $I^{(n)}_{i_1\cdots i_n;i_1\cdots i_n}$ belonging to a particular partition of $n$ with at most $3$ parts are always equal. Accordingly, we can say that the partition of $n$ with at most $3$ parts is uniquely determine the component of tensor $I^{(n)}_{i_1\cdots i_n;i_1\cdots i_n}$ of rank $n$. Another useful property is that the components of $I^{(n)}_{i_1\cdots i_n;i_1\cdots i_n}$ vanish if exactly one or two of $q$, $r$ and $s$ are odd. This property can be seen from invariance under $180^\circ$ rotation of the lab-fixed frame about one of the coordinate axes. For example, the rotation matrix about the $z$-axis is $\mathrm{diag}(-1,-1,1)$ and this rotation requires that $q+r$ be even to have $I(q,r,s)\neq 0$. Likewise $q+s$ and $r+s$ must be even. Briefly, $I(q,r,s)$ can be nonzero (and indeed is, as we will calculate below) only if $q$, $r$, $s$ are all odd or all even. As $n=q+r+s$, this is the same as requiring that $q$, $r$, $s$ have the same parity as $n$.

Interestingly, the number of distinct nonzero components of $I(q,r,s)$ is equal to the number of the Young frames that represent the complete set of linearly independent isotropic tensors \cite{Smith1968,Andrews1977}. 
\item Right-hand side: when we observe that the isotropic tensors $f_r^{(n)}$ and $g_\alpha^{(n)}$ transform among themselves under any permutation of indices, the obtained expressions of independent coefficients only depend on how the indices of $I(q,r,s)$ are partitioned. In other words, for any given partition $n=q+r+s$, the expressions on the right-hand side of Equation~\ref{main2} are the same.
%
\end{enumerate}
In summary, all $I(q,r,s)$ of a given partition $n=q+r+s$ are equal to each other and the equations belonging to the given partition are exactly the same. On the other hand, the number of partitions that provide a nonzero component of $I(q,r,s)$ is $c_n$, the number of independent coefficients on the right-hand side of Equation~\ref{main2}. This tells us that we have the same number of independent equations as variables if we select one equation for each partition $n=q+r+s$ into odd parts (if $n$ is odd).


For the purpose of finding equations for independent coefficients, we have to compute $I(q,r,s)$. In the $z$-$x$-$z$ convention, the direction cosines are parametrized by Euler angles as
\begin{equation}\label{directioncosine}
l=\pmatrix{C_\psi C_\phi-C_\theta S_\phi S_\psi &C_\psi S_\phi+C_\theta C_\phi S_\psi &S_\psi S_\theta\cr
-S_\psi C_\phi-C_\theta S_\phi C_\psi&- S_\psi S_\phi+ C_\theta C_\phi C_\psi& C_\psi S_\theta\cr
S_\theta S_\phi&-S_\theta C_\phi & C_\theta
},
\end{equation}
where $C_\psi=\cos\psi$, $S_\psi=\sin\psi$ and so forth.
Averaging is achieved by Equation~\ref{integration} together with \ref{directioncosine}. Our interest is only in odd rank and as we showed before the powers $q$, $r$, $s$ are all odd for odd rank $n$.

In keeping with a desire to avoid the upper-left $2\times 2$ block as much as possible, we note that $I(q,r,s)=-\langle l_{xx}^ql_{zy}^rl_{yz}^s\rangle$ for odd $n$, as follows from invariance of the average under rotation of the lab-fixed frame by $R$ from Equation~\ref{matrixR}, and it is the latter expression that we explicitly compute.

Recall the elementary trigonometric integrals~\cite{NIST}
\begin{equation}
\int_0^{2\pi}\mathrm{d} x \,\sin^i x\cos^j x=\cases{2\pi\frac{(i-1)!!(j-1)!!}{(i+n)!!}& $i$ and $j$ even\\0 & otherwise}
\end{equation}
\begin{equation}
\int_0^{\pi}\mathrm{d} x \,\sin^i x\cos^j x=2\frac{(i-1)!!(j-1)!!}{(i+j)!!}\qquad \mbox{$i$ odd, $j$ even.}
\end{equation}
These together with Equation~\ref{integration} yield
\begin{equation}
\label{3powodd}
I(q,r,s)=\frac{(r+s)!!}{(q+r)!!(q+s)!!}\sum_{i=0}^{(q-1)/2}\pmatrix{q\cr 2i+1}\frac{[(q-2i-2)!!]^3(2i+r)!!(2i+s)!!}{(q+r+s-2i)!!}.
\end{equation}
In particular,
\begin{equation}\label{2powodd}
I(1,r,s)=\frac{r!!s!!(r+s)!!}{(r+1)!!(s+1)!!(r+s+1)!!}=\pmatrix{\frac r2&\frac s2&\frac{r+s}2\cr\frac12&-\frac12&0}^2
\end{equation}
and
\begin{equation}\label{1powodd}
I(1,1,s)=\frac1{2(s+2)}.
\end{equation}
The $3j$ symbol in Equation~\ref{2powodd} relates to rotational averages of Wigner $D$-matrix elements \cite[Chapter 4]{Edmonds1957}, though we will not pursue this interesting connection.

{\it Example 1.} $\mathbf{M}'^{(5)}$. There are $N_5=10$ different linearly dependent isotropic tensors for rank $n=5$, composed of 10 different epsilon tensors multiplied by a Kronecker delta symbol, which is the only isotropic tensor for rank $n=2$, namely
\begin{eqnarray}
&f_1^{(5)}=\epsilon_{i_1i_2i_3}\delta_{i_4i_5},\quad f_6^{(5)}=\epsilon_{i_1i_4i_5}\delta_{i_2i_3},\nonumber\\
&f_2^{(5)}=\epsilon_{i_1i_2i_4}\delta_{i_3i_5},\quad f_7^{(5)}=\epsilon_{i_2i_3i_4}\delta_{i_1i_5},\nonumber\\
&f_3^{(5)}=\epsilon_{i_1i_2i_5}\delta_{i_3i_4},\quad f_8^{(5)}=\epsilon_{i_2i_3i_5}\delta_{i_1i_4},\nonumber\\
&f_4^{(5)}=\epsilon_{i_1i_3i_4}\delta_{i_2i_5},\quad f_9^{(5)}=\epsilon_{i_2i_4i_5}\delta_{i_1i_3},\nonumber\\
&f_5^{(5)}=\epsilon_{i_1i_3i_5}\delta_{i_2i_4},\quad f_{10}^{(5)}=\epsilon_{i_3i_4i_5}\delta_{i_1i_2}
\end{eqnarray}
and $\mathbf{M}'^{(5)}$ is a $10\times 10$ scalar matrix. That is, $\mathbf{M}'^{(5)}=a\mathbf{E}$ where $\mathbf{E}$ is the $10\times 10$ unit matrix and $a$ is the only coefficient that needs to be determined. Consequently, the rotational average of direction cosines can be written as 
\begin{equation}\label{fifth0}
I^{(5)}_{i_1\cdots i_5;\lambda_1\cdots\lambda_5}=a\sum_{r,\alpha}f_r^{(5)}g_\alpha^{(5)}
\end{equation}
where $r$ and $\alpha$ range from $1$ to $10$ and its diagonal term is
\begin{equation}
I^{(5)}_{i_1\cdots i_5;i_1\cdots i_5}=a(\epsilon_{i_1i_2i_3}\delta_{i_4i_5}\epsilon_{i_1i_2i_3}\delta_{i_4i_5}+\epsilon_{i_1i_2i_4}\delta_{i_3i_5}\epsilon_{i_1i_2i_4}\delta_{i_3i_5}+\ldots).
\end{equation}
The diagonal term $I(1,1,3)=1/10$ according to Equation~\ref{1powodd} and the resulting equation is
\begin{equation}
\frac{1}{10}=I(1,1,3)=I^{(5)}_{xyzzz;xyzzz}=3a.
\end{equation}
The coefficient $a$ can be found as $1/30$. This solution is consistent with the result obtained by others \cite{Power1974,Tinoco1975,Andrews1977}. 

{\it Example 2.} $\mathbf{M}'^{(7)}$. There are $N_7=105$ linearly dependent isotropic tensors of rank $n=7$. These isotropic tensors can be classified into 35 equally divided groups. Each group has the same epsilon tensor but different Kronecker deltas. For example, the first and last groups are
\begin{eqnarray}
&f_1^{(7)}=\epsilon_{i_1i_2i_3}\delta_{i_4i_5}\delta_{i_6i_7}, \qquad &f_{103}^{(7)}=\epsilon_{i_5i_6i_7}\delta_{i_1i_2}\delta_{i_3i_4},\nonumber\\
&f_2^{(7)}=\epsilon_{i_1i_2i_3}\delta_{i_4i_6}\delta_{i_5i_7}, \qquad &f_{104}^{(7)}=\epsilon_{i_5i_6i_7}\delta_{i_1i_3}\delta_{i_2i_4},\nonumber\\
&f_3^{(7)}=\epsilon_{i_1i_2i_3}\delta_{i_4i_7}\delta_{i_5i_6}, \qquad \mbox{and}\qquad &f_{105}^{(7)}=\epsilon_{i_5i_6i_7}\delta_{i_1i_4}\delta_{i_2i_3}.
\end{eqnarray}
Each group has the same structure for Kronecker deltas. Particularly, the product of Kronecker deltas $f_1^{(4)}$ appearing in the first member of each group has indices in ascending order. The second and third isotropic tensors are obtained by certain permutations of indices of $f^{(4)}_1$. The permutations are the same for all groups.

The matrix $\mathbf{M}'^{(7)}$ in the set of 105 linearly dependent isotropic tensors has a block diagonal form. Each block $\mathbf{A}^{(4)}$ is of dimension $3\times 3$, and has the same structure as $\mathbf{M}'^{(4)}$ given by D.~L.~Andrews \cite{Andrews1977} as
\begin{equation}
\mathbf{A}^{(4)}=
\pmatrix{
a & b & b\cr
b & a & b\cr
b & b & a
},
\end{equation}
where $a$ and $b$ are independent coefficients. The two admissible partitions of $7$ are $1,1,5$ and $1,3,3$. The corresponding components of $I^{(7)}$ are found to be $I(1,1,5)=1/14$ and $I(1,3,3)=9/140$ according to formulas \ref{1powodd} and \ref{2powodd}.
The coupled equations for $a$ and $b$ are
\begin{eqnarray}
\frac{1}{14}&=I(1,1,5)=I^{(7)}_{xyzzzzz;xyzzzzz}=15a+30b,\nonumber\\
\frac{9}{140}&=I(1,3,3)=I^{(7)}_{xyyyzzz;xyyyzzz}=9a,
\end{eqnarray}
where the expressions on the right-hand side follow from Equation~\ref{main2}.
The unique solution is $(a,b)=(6/840,-1/840)$, in agreement with the result of D.~L.~Andrews et al.\ \cite{Andrews1977}.

\section{Rotational average of a ninth-rank tensor}
Based on the previous discussion, the matrix $\mathbf{M}'^{(9)}$ for ninth-rank tensors has a block diagonal form in the linearly dependent set which consists of products of 84 epsilon tensors and 15 isotropic tensors of rank 6. The isotropic tensors of rank 6 are given by D.~L.~Andrews et al. \cite{Andrews1977}, and we use the same ordering as they did to enumerate them. Then the linearly dependent isotropic tensors of rank 9 are $\epsilon_{i_1i_2i_3}f^{(6)}_r$, $\epsilon_{i_1i_2i_4}f^{(6)}_r$, ... ,$\epsilon_{i_7i_8i_9}f^{(6)}_r$, where $f^{(6)}_r$ is the $r$th isotropic tensor of rank 6 and indices are composed of unused indices in the corresponding epsilon tensor. Therefore,
\begin{equation}
\mathbf{M}'^{(9)}=\mathbf{E}\otimes \mathbf{A}^{(6)}
\end{equation}
where
\begin{equation}\label{aaa}
\mathbf{A}^{(6)}=
\pmatrix{
a & b & b & b & c & c & b & c & c & c & c & b & c & c & b\cr
b & a & b & c & b & c & c & c & b & b & c & c & c & b & c\cr
b & b & a & c & c & b & c & b & c & c & b & c & b & c & c\cr
b & c & c & a & b & b & b & c & c & c & b & c & c & b & c\cr
c & b & c & b & a & b & c & b & c & b & c & c & c & c & b\cr
c & c & b & b & b & a & c & c & b & c & c & b & b & c & c\cr
b & c & c & b & c & c & a & b & b & b & c & c & b & c & c\cr
c & c & b & c & b & c & b & a & b & c & b & c & c & c & b\cr
c & b & c & c & c & b & b & b & a & c & c & b & c & b & c\cr
c & b & c & c & b & c & b & c & c & a & b & b & b & c & c\cr
c & c & b & b & c & c & c & b & c & b & a & b & c & b & c\cr
b & c & c & c & c & b & c & c & b & b & b & a & c & c & b\cr
c & c & b & c & c & b & b & c & c & b & c & c & a & b & b\cr
c & b & c & b & c & c & c & c & b & c & b & c & b & a & b\cr
b & c & c & c & b & c & c & b & c & c & c & b & b & b & a\cr
}
\end{equation}
has the same structure as $\mathbf{M}'^{(6)}$ given in ref.~\cite{Andrews1977}.
Here, $\mathbf{E}$ is the unit matrix of dimension $84\times 84$. There are three independent coefficients which we denote $a$, $b$ and $c$. The system of linear equations for these coefficients can be found by computing $I(1,1,7)$, $I(1,3,5)$ and $I(3,3,3)$ and using Equation~\ref{main2}. Resulting equations are
\begin{eqnarray}
\frac{1}{18}&=105a+630b+840c,\nonumber\\
\frac{1}{21}&=45a+90b,\nonumber\\
\frac{19}{420}&=27a
\end{eqnarray}
with the solution 
\begin{equation}\label{result}
a= \frac{38}{22680},\qquad
b= -\frac{7}{22680},\qquad
c= \frac{2}{22680}.
\end{equation}
Substituting the obtained numbers Equation \ref{result} into the matrix $\mathbf{A}^{(6)}$ given by Equation \ref{aaa} and assembling 84 copies of $\mathbf{A}^{(6)}$ into a block diagonal matrix we find the rotational average $I^{(9)}$ in the linearly dependent set of isotropic tensors.

\section{Rotational average of an eleventh-rank tensor}
In the case of eleventh-rank tensors, there are $N_{11}=17325$ linearly dependent isotropic tensors which can be divided into 165 groups. Each group has 105 isotropic tensors determined by eighth-rank isotropic tensors. As we did in the case of $n=9$, the matrix $\mathbf{M}'^{(11)}$ can be written as $\mathbf{M}'^{(11)}=\mathbf{E}\otimes \mathbf{A}^{(8)}$. Here, $\mathbf{A}^{(8)}$ has the same structure as $\mathbf{M}'^{(8)}$ given by D.~L.~Andrews et al.\ \cite{Andrews1981}, and $\mathbf{E}$ is the unit matrix of dimension $165\times 165$. There are four independent coefficients $a$, $b$, $c$ and $d$ in the matrix $\mathbf{M}'^{(11)}$; here we cast the coefficients $A$, $B$, $C$ and $D$ in $\mathbf{M}'^{(8)}$ in ref.~\cite{Andrews1981} into lower case.
The calculation procedure is the same as we did for the ranks $n=5,7,9$ and straightforward. However, we perform the calculation via computer since it is so lengthy. As a result, we obtain the equations for independent coefficients as follows:
\begin{eqnarray}
\frac{1}{22}&=945a+11340b+11340c+30240d,\nonumber\\
\frac{5}{132}&=315a+1890b+2520d,\nonumber\\
\frac{25}{693}&=225a+900b+900c,\nonumber\\
\frac{97}{2772}&=135a+270b
\end{eqnarray}
with the solution
\begin{eqnarray}
a&=\frac{548}{1496880}, \qquad b=-\frac{80}{1496880},\nonumber\\
c&=\frac{3}{1496880}, \qquad d=\frac{14}{1496880}.
\end{eqnarray}

\section{Concluding remark}
We present a new method for three-dimensional rotational averages of odd-rank tensors. The method is applied to low-rank tensors $n=5,7$ as an example and also applied to ninth- and eleventh-rank tensors that were not known before in explicit form. The results of our method $I^{(n)}_{i_1\cdots i_n;\lambda_1\cdots\lambda_n}$ (rotational average of odd-rank tensors) are expressed in block diagonal form in the overcomplete set of isotropic tensors. Fortunately, the number of independent coefficients that determine $I^{(n)}_{i_1\cdots i_n;\lambda_1\cdots\lambda_n}$ is just three and four for ninth- and eleventh-rank tensors, respectively. These coefficients are found in the present paper. The obtained three-dimensional rotational averages of odd-rank tensors can be used for calculation in various type of nonlinear spectroscopy in optically active medium.

\ack
We are grateful to the Air Force Office of Scientific Research (Award No. FA9550-18-1-0141), the Office of Naval Research (Award No. N00014-16-1-3054), and the Robert A. Welch Foundation (Grant No. A-1261).

\section*{References}
\bibliographystyle{iopart-num}
\bibliography{nomzui}   

\providecommand{\newblock}{}
\begin{thebibliography}{10}
\expandafter\ifx\csname url\endcsname\relax
  \def\url#1{{\tt #1}}\fi
\expandafter\ifx\csname urlprefix\endcsname\relax\def\urlprefix{URL }\fi
\providecommand{\eprint}[2][]{\url{#2}}

\bibitem{Craig1998}
Craig D and Thirunamachandran T 1998 {\em Molecular Quantum Electrodynamics: An
  Introduction to Radiation-molecule Interactions\/} Dover Books on Chemistry
  Series (Dover Publications)

\bibitem{Smith1968}
Smith G~F 1968 {\em Tensor, N. S.\/} {\bf 19} 79--88

\bibitem{Boyle1970}
Boyle L~L 1970 {\em Int. J. Quantum Chem.\/} {\bf 4} 413--425

\bibitem{Boyle1971}
Boyle L~L and Matthews P~S~C 1971 {\em Int. J. Quantum Chem.\/} {\bf 5}
  381--386

\bibitem{Andrews1977}
Andrews D~L and Thirunamachandran T 1977 {\em J. Chem. Phys.\/} {\bf 67}
  5026--5033

\bibitem{Andrews1981}
Andrews D~L and Ghoul W~A 1981 {\em J. Phys. A: Math. Gen.\/} {\bf 14} 1281

\bibitem{Wagniere1982}
Wagni{\`e}re G 1982 {\em J. Chem. Phys.\/} {\bf 76} 473--480

\bibitem{Andrews1989}
Andrews D~L and Blake N~P 1989 {\em J. Phys. A: Math. Gen.\/} {\bf 22} 49

\bibitem{Smith2011}
Smith S~N~A and Andrews D~L 2011 {\em J. Phys. A: Math. Theor.\/} {\bf 44}
  395001

\bibitem{Man2014}
Man P~P 2014 {\em Concepts Magn. Reson., Part A\/} {\bf 42} 197--244

\bibitem{Friese2014}
Friese D~H, Beerepoot M~T~P and Ruud K 2014 {\em J. Chem. Phys.\/} {\bf 141}
  204103

\bibitem{Ford2018}
Ford J~S and Andrews D~L 2018 {\em J. Phys. Chem. A\/} {\bf 122} 563--573

\bibitem{Bjarnason1980}
Bjarnason J~O, Andersen H~C and Hudson B~S 1980 {\em J. Chem. Phys.\/} {\bf 72}
  4132--4140

\bibitem{Oudar1982}
Oudar J, Minot C and Garetz B~A 1982 {\em J. Chem. Phys.\/} {\bf 76} 2227--2237

\bibitem{Koroteev1995}
Koroteev N~I 1995 {\em Biospectroscopy\/} {\bf 1} 341--350

\bibitem{Zheltikov1999}
Zheltikov A~M and Naumov A~N 1999 {\em Quantum Electron.\/} {\bf 29} 607

\bibitem{Hiramatsu2015}
Hiramatsu K, Leproux P, Couderc V, Nagata T and Kano H 2015 {\em Opt. Lett.\/}
  {\bf 40} 4170--4173

\bibitem{Begzjav2018}
Begzjav T~K, Zhang Z, Scully M~O and Agarwal G~S  TBP

\bibitem{Weyl1939book}
Weyl H 1939 {\em The Classical Groups: Their Invariants and Representations\/}
  Princeton mathematical series (Princeton University Press)

\bibitem{NIST}
{\it NIST Digital Library of Mathematical Functions} http://dlmf.nist.gov/,
  Release 1.0.20 of 2018-09-15 f.~W.~J. Olver, A.~B. {Olde Daalhuis}, D.~W.
  Lozier, B.~I. Schneider, R.~F. Boisvert, C.~W. Clark, B.~R. Miller and B.~V.
  Saunders, eds. (see Eqs.~5.12.1 and 5.12.2)

\bibitem{Edmonds1957}
Edmonds A~R 1957 {\em Angular Momentum in Quantum Mechanics\/} Investigations
  in physics (Princeton University Press)

\bibitem{Power1974}
Power E~A and Thirunamachandran T 1974 {\em J. Chem. Phys.\/} {\bf 60}
  3695--3701

\bibitem{Tinoco1975}
Tinoco I 1975 {\em J. Chem. Phys.\/} {\bf 62} 1006--1009

\end{thebibliography}

\end{document}